\documentclass[twocolumn]{article}
\usepackage{setspace}
\usepackage{graphicx}
\usepackage[super]{natbib}
\usepackage{amssymb, latexsym}
\usepackage{amsmath} 
\usepackage{amsthm}
\usepackage{bm}
\usepackage[mathscr]{eucal}
\usepackage{enumerate}
\usepackage{multirow} 
\usepackage[center]{caption}
\usepackage{musicography}
\usepackage{float}

\usepackage{geometry}

\usepackage{authblk}
\usepackage{abstract}

\begin{document}
	
	\title{Neural Synchronization of Music Large-Scale Form}
	
	\author{Lenz Hartmann and Rolf Bader}
	\affil{ Institute of Systematic Musicology\\ University of
		Hamburg\\ Neue Rabenstr. 13, 20354 Hamburg, Germany\\
	}
	\date{\today}
	
\twocolumn
[
\begin{@twocolumnfalse}

	\maketitle
	\begin{abstract}
Music large-scale form, the structure of musical units ranging over several bars, are studied using EEG measurements of 25 participants listened to the first four minutes of a piece of electronic dance music (EDM).
Grand-averages of event-related potentials (ERPs) calculated for all electrodes show dynamics in phase synchronization between different brain regions. Here local maxima of the perceptual parameters correspond to strong synchronization, which culminate at time points, where musical large-scale form boundaries were perceptually expected. 
Significant differences between local maxima and minima were found, using a Paired Samples t-test, showing global neural synchronization between different brain regions most strongly in the gamma-band EEG frequency range. Such synchronization increases before musical large-scale form boundaries, and decreases afterwards, therefore representing musical large-scale form perception. 
		\end{abstract}
\end{@twocolumnfalse}
]

\vspace{1cm}
\renewcommand \thesection{\Roman{section}}
\renewcommand \thesubsection{\Alph{subsection}} 
\renewcommand \thesubsubsection{\arabic{subsubsection}} 

\section{Introduction}
\subsection{Music Large-scale form}
  Music large-scale form as investigated in this paper is meant to be the overall structure of a piece of music. The verse and chorus concatenation in a song, the sonata form of classical music, or the continuous night-long tension built-up and decay in Techno, House, or Electronic Dance Music (EDM) are all examples of such large-scale forms. So in the hierarchical structuring of music form is the perceptual highest level of grouping.

Although this paper is investigating large-scale forms of EDM, all forms are closely related to the creation and release of perceived tension and expectations. In terms of music theory this was prominently been discussed already by Hugo Riemann in 1895 when suggesting a cadence to consist of functions, therefore introducing functional harmony. There he used the Hegel terms of thetic-antithetic-synthetic (These-Antithese-Synthese) for the chord progression I-IV-I-V-I, finding the V to have a maximum tension, demanding the tension release in the following I \cite{Riemann.1976}\cite{Bader.2009}. 

While listening to music, the incoming flow of auditory information is perceptually segregated \cite{Bregman.1990} and organized onto different levels \cite{Lerdahl.1990} by the principles of Gestalt psychology \cite{Wagemans.2012,Wagemans.2012b,Deutsch.2013}. These principles apply to the formation of primitive auditory objects like pitches or chords \cite{Stumpf.1965}, as well as to the organization of phrases and greater passages in pieces of music \cite{Lerdahl.1990,Koelsch.2013}. These structural aspects on the highest level of musical organization ranging over several bars as a combination of all elements that constitute a piece of music, like pitch, rhythm and timbre, are here referred to as musical form.

In the early 1930s, music theory was much influenced by energetics (Energetiker) \cite{Bader.2013,Burnham.2006}. In 1931 the music theory of Ernst Kurth described music as an interplay of potential and kinetic energy, where tension is characterized by high potential energy that will be converted into kinetic energy, therefore lowering the potential energy \cite{Kurth.1931}. According to Rothfarb this idea was supported by Arnold Schering \cite{Rothfarb.2006}, for whom the essence of music was alternating phases of tension and release. More modern theories like the Generative Theory of Tonal Music (GTTM) incorporate tension as well, namely between strong metrical events \cite{Lerdahl.1990}. In general, Sch"onberg characterized the musical form by the contrasts of its subsidiary parts, or in his own words: “Larger forms develop through the generating power of contrasts.” \cite[p. 178]{Schoenberg.1967}. Due to this contrasting elements in different form parts, the use of letters as formal representation of the different form parts has been established, like ABA, ABA’ or ABAC, where A denotes the first part, B a second part and A’ a variation of the first part A \cite{Burnham.2006,Neuhaus.2013}.

\subsection{Music form in 'no-score' music}
As most theories have been investigating form extracting a musical score. These theories are not applicable to modern dance music like House, Techno, or EDM in a straightforward way, as these musical styles most often have no scores at all. Still such music has form, an overall structure, and like most musical styles this form follows a certain typical schema \cite{Snoman.2009}. This schema is based on the metrical structure of a piece in such a way that instrumentation and looped parts or grooves change only at metrically strong positions, namely after a multiple of 16 or 32 bars \cite{Lerdahl.1990,Snoman.2009}. These highest-level musical events are constitutive for the musical form, since they determine the boundaries of the different form parts \cite{Deliege.2014}.

\subsection{Musical schema cause expectations}
Perceptually, these schema \cite{Huron.2006,Huron.2012}, or normative archetypes \cite{Meyer.1956}, give rise to schematic or high-level structural expectations \cite{Meyer.1956,Pearce.2012,Huron.2006,Huron.2012}. Since House, Techno, or EDM music rely on repeated or looped parts or grooves, based on a 16- or 32-bar schema, the time points of high-level musical part boundaries are predictable, and so expectations and tension emerge while listening. When the predicted event occurs, and so expectation is fulfilled, tension releases \cite{Huron.2006,Snoman.2009,Friston.2013}. In the context to Dance music, and with respect to Ernst Kurth's alternating potential and kinetic energies, these releases motivate movements of listeners dancing \cite{Solberg.2016,Witek.2016}.

\begin{figure*}[t]
		\includegraphics[width=1\textwidth]{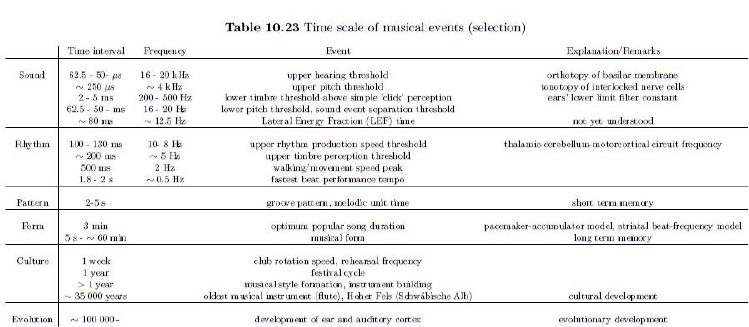}
		\caption{Time scale of musical events (selection). \cite[p. 324]{Bader.2013}. Reprinted with permission.}
	\label{fig:Bader_eventscale}
\end{figure*}

Expectations are often found to emerge from learning processes, and in particular from statistical learning \cite{Huron.2012,Friston.2013}. Since musical events happen on different time-scales, from milliseconds to life-time spans [Figure \ref{fig:Bader_eventscale}], also expectations built-up for musical events on different time-scales.

Zanto showed gamma-band synchronization in response to perturbed auditory pulse sequences in an EGG-study with eight subjects \cite{Zanto.2005}. Evoked gamma-band activity (phase-locked to stimulus) was measured at metrical positions where an auditory pulse actually appeared, and induced gamma-band activity (time-locked to stimulus) was measured at the sites where an auditory pulse should have occurred, but actually was not presented physically.

In a more complex context Jones et al. investigated the influence of the precise temporal placement of a target tone on the judgments of a pitch-comparison task \cite{Jones.2002}. It was shown that the judgments were more accurate the more precise the target tone was placed on a metrical strong position. The authors interpreted the results in that way that attention is not equally distributed over time, and in general support the idea of Seashore that attention may be periodic \cite{Seashore.1938}.

Krumhansl et al. were able to show culturally influenced learning and expectations in a probe-tone experiment with tone-scales \cite{Krumhansl.1979}. Subjects were asked to judge how good a certain probe-tone fits to a previously heard tone-scale from the pool of all 24 major and minor scales and tones. The analysis of the presented intervals for major and minor scales, the so-called tone-profiles, show that octave, fifth, and major or minor third, are rated highest. These results support the hypothesis of cultural learning of tonal relations within a certain tonal system, and as such was also modeled using an artificial neural network \cite{Leman.1995,Bader.2013}.

\subsection{Timing representation in the brain}
Less empirical research was performed on how music large-scale form is processed by the human brain. However, there are expectation models independent of musical interpretations, such as the pacemaker-accumulator model\cite{Gibbon.1977,Gibbon.1984} or the synchronization model by \cite{Buhusi.2005,Buhusi.2009}.

Time perception, like music, is organized on different time-scales: the circadian timing, the interval timing, and the millisecond timing. The circadian timing is linked to the 24h sleep-wake cycle and appetite, and its corresponding neural timer is located in the suprachiasmatic nucleus of the hypothalamus \cite{Reppert.2002,Buhusi.2005}. Millisecond timing is crucial for motor control and speech generation, and is discussed to be neurally represented in the cerebellum \cite{Buhusi.2005}.

Yet the third time-span is in a seconds-to-minutes range, and referred to as interval timing. Interval timing is found to be cognitively controlled, and involved in decision-making and conscious time estimation. Interval timing is characterized by the scalar property that has been found in several timing reproduction task in a wide variety of animals’ species. By asking subjects to reproduce a given duration, the responses are normally distributed around that duration, and the distribution proportional to the duration \cite{Gibbon.1977,Gibbon.1984,Buhusi.2005}. A model of timing therefore has to reproduce this scalar property.

The traditional explanation of the scalar property is the pacemaker-accumulator model \cite{Gibbon.1977,Gibbon.1984} as a scalar expectancy theory. Here, the internal clock is represented by a pacemaker that sends pulses to an accumulator which stores the pulses until a certain feedback or reward occurs. The number of stored pulses represents the time-span, and is stored in a reference memory. When reproducing the time-span, the current number of pulses is compared to the number in the reference memory. The scalar property is explained by the proportionality between the accumulation error and the criterion duration \cite{Gibbon.1977,Buhusi.2005}.

Buhushi postulated a neural model based on a review of literature on neural mechanisms, where interval timing is described by a coincidence-detection model \cite{Buhusi.2005}. Cortical oscillations modulate neural activity in the striatum, which acts as a coincidence detector of the phases of these cortical oscillations. As more and more cortical oscillations synchronize, striatal neurons monitor this by an increasing activity and dopamine release, thereby having the scalar property \cite{Holson.1996,Buhusi.2005}.

\subsection{Neural synchronization}
Concerning perception, the brain is considered as a Helmholtz machine \cite{Dayan.1995}, a self-organizing system \cite{Kelso.1997,Friston.2006,Friston.2007,Bader.2013,Friston.2013,Freeman.2013,Freeman.2015,GuevaraErra.2017} “that actively constructs predictions and or explanations for sensory input using internal or generative models.” \cite[p. 44]{Friston.2013}.

In that context, synchronization is seen as a far-ranging principle used by neurons or neural ensembles to code and process information \cite{Salinas.2001,Fries.2009,GuevaraErra.2017,Nowak.2017,Palacios.2019}. While a total synchronization of all neurons is associated with an epileptic seizure \cite{Pastor.2012,Freeman.2013,Jiruska.2013}, partly synchronization of different neurons or neural ensembles, locally or globally distributed over different brain regions, is associated with cognitive and perceptual processes \cite{Bhattacharya.2001,Mima.2001,Melloni.2007,Perez.2017}.

Especially large-scale synchronization of cortical neurons are interesting, since they are associated with a bunch of cortical and perceptual processes like gestalt perception \cite{Gray.1989,Tallon.1995,TallonBaudry.1999,Rodriguez.1999,Engel.2001,Engel.2001b}, timing and expectation \cite{Buhusi.2005}, attention\cite{Womelsdorf.2007,Fries.2009,Nikolic.2013}, consciousness \cite{Dehaene.2011b,Owen.2019,Baars.2006}, or motor functions and entrainment of motor neurons by auditory neurons \cite{Thaut.2015}

In the context of music perception, neural synchronization in various frequency bands and domains has been found in various perceptual and cognitive tasks. An early approach of studying communication between different brain regions in relation to music perception was proposed by \cite{Petsche.1988}. In their paper “The EEG: An Adequate Method to Concretize Brain Processes Elicited by Music” in an experiment with 75 subjects they examined  if several EEG parameters (location, power, frequency, and coherence) differ between groups of musicians and non-musicians with respect to different musical tasks. They emphasize that coherence in perceptional tasks “reflects different degrees of functional coordination of two adjacent brain regions or the two hemispheres” \cite[p. 133]{Petsche.1988}. In a following experiment Janata showed that neural coherence, and to a smaller extend neural amplitude, can predict if a musical context is completed, that is, if the context generated expectation is fulfilled or not \cite{Janata.1993}. Coherence is not synchronization, but still it describes simultaneous and frequency depending activity of different brain regions. As it was common practice in these times, mostly established by technical restrictions, mainly frequencies slower than 30Hz were in the focus of interest.

The introduction of gamma-frequencies (faster than 30Hz, mostly specified between 30Hz and 80Hz) as a (firstly visual) perception related frequency-band in the mid to late 1990th \cite{TallonBaudry.1999} extended the experimental focus to faster frequencies. Bhattacharya examined long-range synchronization in musicians and non-musicians in the gamma-band spectrum \cite{Bhattacharya.2001}. They found that gamma-band synchrony over distributed cortical areas were significantly higher in musicians than non-musicians while listening to music. No differences were found in control conditions (listening to text and at rest). Another interesting study concerning synchronization in the gamma-band was realized by Zanto et al. \cite{Zanto.2005}. In an EEG experiment they examined gamma-band activity in the averaged EEG activity of eight subjects as they listened to isochronous pure-tone sequences with embedded temporal perturbations. They found that induced (not phase-locked) gamma activity was enhanced at the occurrence of tone onsets, while evoked (phase-locked) gamma-band activity was observed after onset. At late perturbations, induced gamma-band activity, peaks precede tone onset, during early perturbations, induced activity following tone onset. The authors interfere that induced gamma-band activity represents metrical expectationm and evoked gamma-band activity represents stimulus perception. Snyder et al. went a step further and examined metrical structures instead of pulse sequences \cite{Snyder.2005}. They could confirm these results, and proposed that the power of evoked gamma-band activity reflects the loudness of the stimuli.

Besides experimental data, different models use mechanisms of neural synchronization to explain phenomena in music perception and physiology. In line with the results concerning the perception of pulse and meter above, Large et al. developed a model to explain beat perception in musical rhythms by synchronization of oscillations in self-organizing neural networks \cite{Large.2015}. Concerning pitch perception in the auditory system, Bader used a coincidence detection model to explain the synchronization of the blurred spike output from the cochlear in the nucleus cochlearis and the trapezoid body found in cat auditory nerves \cite{Bader.2018} \cite{Joris.1994,Joris.1994b}. It shows that one single neuron could detect pitches up to 300Hz by neural coincidence.

This paper therefore hypothesis that when listening to a piece of music, the different hierarchical organized parts of the piece are perceptually integrated into a high-level Gestalt, so that musical form emerges over time. At the same time, expectations are formed towards the high-level musical events or cue point that determine the boarders of the individual parts. Since both processes are represented by dynamics of neural synchronization, we hypothesis that large-scale neural synchronization increases before a musical event occurs and decrease afterwards, with a maximum synchronization at the time of the musical event. 

With EDM music the raise and fall of perceived tension is a major element in each performance. During a DJ set lasting maybe a whole night, over and over again tension is increased most often by an increase of instrumentation density, brightness, or amplitude raise, which then is released when the bass drum comes in, in a so-called four-to-the-floor beat, meaning the bass drum is played at each quarter note. Therefore EDM music is used in this investigation.

\section{Method}
In order to test the hypothesis whether the synchronization at the time of the musical events is higher than at previous synchronization minima, we conducted an EEG experiment with 25 participants listening to a piece of EDM music. Then the synchronization between different electrodes of the averaged brain activity of the test person was calculated in successive time intervals.
Subsequently, the differences between the strength of synchronization at the time of the musical events and the preceding local synchronization minima were tested with a paired sample t test.


\begin{figure}[H]
	\centering
	\includegraphics[width=1\linewidth]{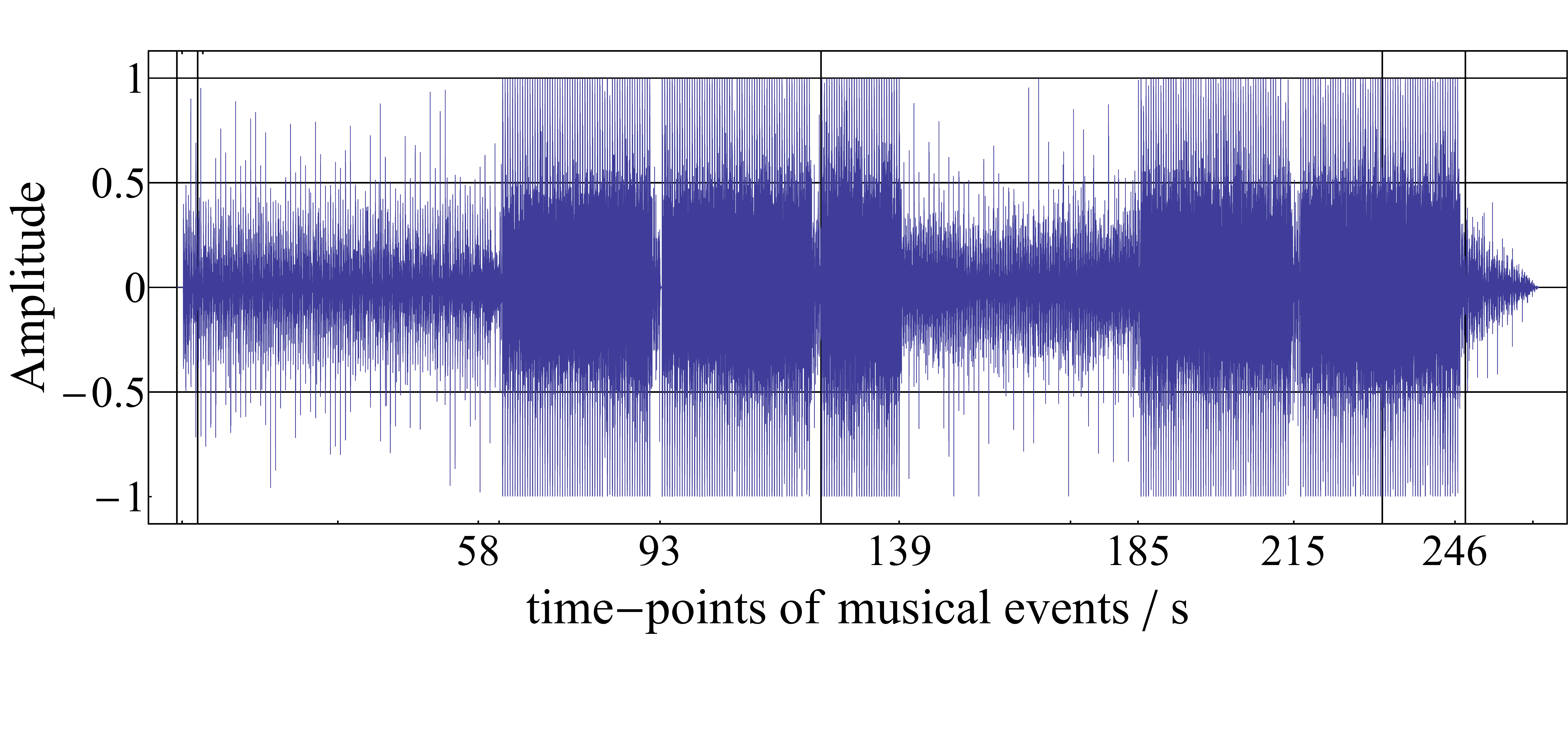}
	\caption{Wave form of the stimulus.}
		\label{fig:Farben_wave}
	\end{figure}

The stimulus (the musical piece) was selected according to various criteria. The piece should follow a clear structure, contain predictable musical events, so elicit expectations, and be accessible to the listener familiar with the genre. These elements are prerequisites in modern dance music, since genres such as Techno or House follow clear compositional structures \cite{Snoman.2009,Solberg.2016}. Also from a practical point of view it is easier to find test persons for these genres than for more complex structured music.

\begin{figure}[H]
	\centering
	\includegraphics[width=1\linewidth]{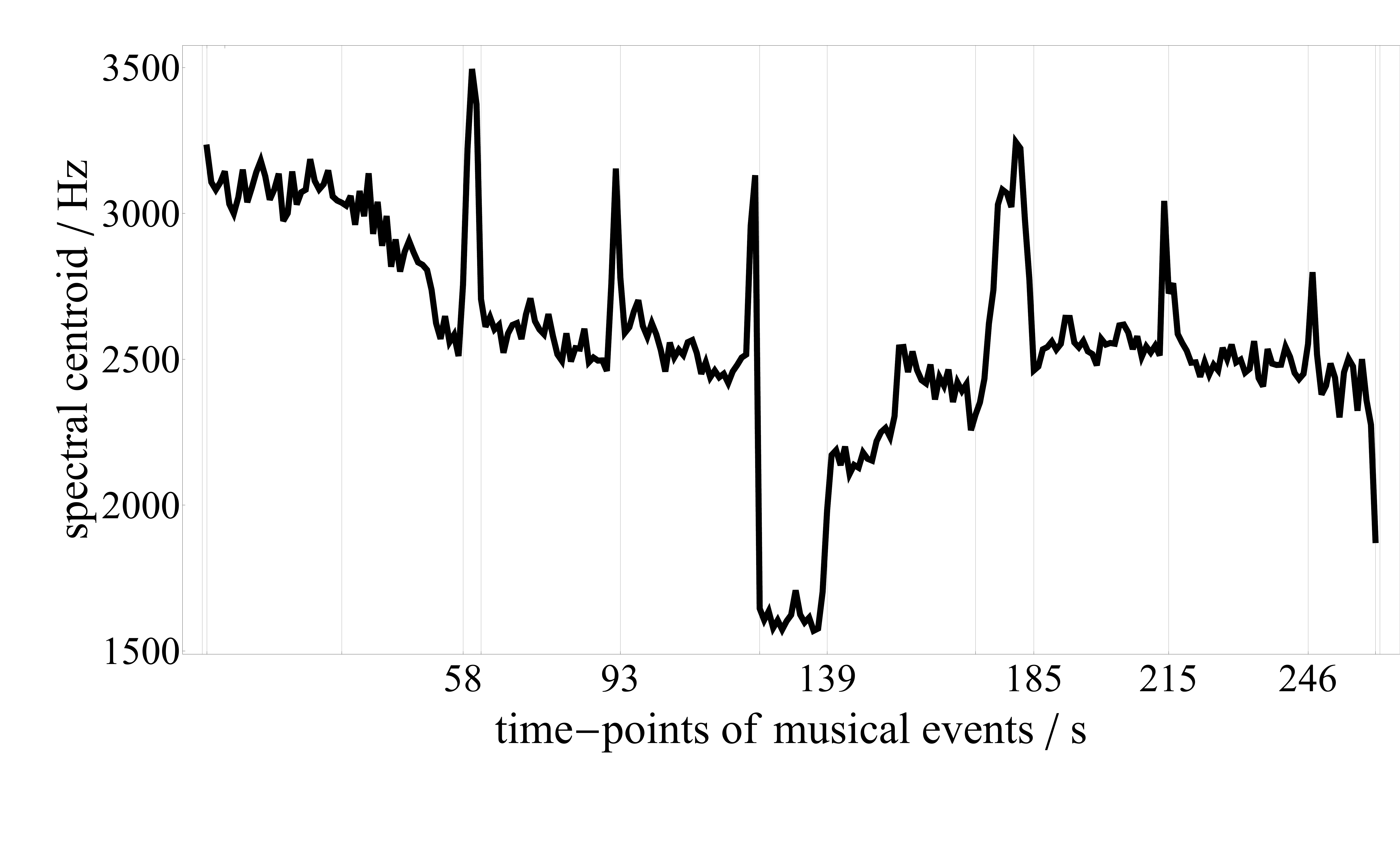}
	\caption{Spectral centroid of the stimulus.}
		\label{fig:Farben_SC}
	\end{figure}

We chose the piece 'Classical Symphony' by Shemian in the remix of Alle Farben because it meets these requirements \footnote{https://www.youtube.com/watch?v=YauWD68O2RE}. We took the first 128 bars from the piece. It follows a 16-bar structure with a 4/4-time signature, where parts with bassline and parts without bassline alternate. The piece has a tempo of 125 beats per minute (bpm). Fig. \ref{fig:Farben_wave} shows the amplitude of the stimulus over time. The differences between parts with low and high amplitude are clearly visible. Fig. \ref{fig:Farben_SC} shows the spectral centroid calculated for subsequent time windows of one second. The short peaks in Fig. \ref{fig:Farben_SC} show fast changes in the spectral midpoint, which are indicators of the important musical events separating the musical form into different parts. Following this reasoning the musical events occur at the 58th, 93rd, 139th, 185th, 215th, and 246th second. The solid vertical lines represent the boarders between the form parts, following a typical Tech-House style sixteen-bars schema. A few extra lines were added to clarify the structure of the stimulus. 

\subsection{Subjects}
Twenty-five subjects, nine women and sixteen men aged between 20 and 32 with a mean of 26.8 participated in this study. During the experiment participants sat half lying half sitting on a recliner in front of a white wall. They were instructed to avoid, but explicitly not to suppress body movements. The eyes were opened and eye movements were not forbidden in order to avoid a behavioral concentration to that goal. After the setup and instructions, the stimulus was presented three times to each subject via a pair of headphones (Sennheiser HD203, frequency range: 18-18.000Hz) at a medium listening volume.

\subsection{Procedure and recordings}
Electroencephalographic (EEG) signals were recorded from 32 electrodes (Electrode-cap Wave-guard 32, ANT-Neuro). All impedances were kept below 10 kOhm. The electrodes were positioned following the 10-20 method of placement \cite{Jasper.1958}, and were referred to the average of all electrodes with a middle forehead ground. The EEG signal was processed by an amplifier (Refa-32, ANT-Neuro) with a sampling frequency of 500Hz. The recorded data were transferred to a computer, where the EEG processing software ASA (V. 4.7.3, ANT-Neuro) was used for recording and pre-processing.

\subsection{Data analysis}
The data analysis consisted of three parts: the pre-processing, the calculation of the synchronization, and the hypothesis testing, whether the strength of the global synchronization differs between the time points of the minima before a musical event and the time points of the musical events.

\subsubsection{Pre-processing}

The recorded data consisted of 75 datasets, 25 subjects $\times$ 3 trials, containing the EEG signal of the 32 electrodes over stimulus length of 4:22 min. The time series of each electrode represented the measured voltage fluctuations over time in $\mu$V. The pre-processing consisted of three steps:

\begin{enumerate}
	\item Artefacts: In a recorded EEG signal all potential fluctuations that are not induced by brain activity are regarded as artefacts. The most common artefacts are induced by eye-blinks, body movements, or transpiration. Eye-blinks were corrected using the ASA Artefact correction feature, which is based on a principle components analysis of the whole signal to separate artefacts from stimulus correlated brain activity \cite{Ille.2002}. Body movements cause muscle artefacts, which can be found in the EEG raw data by visual inspection. Since these artefacts are strongly not-linear, and so hard to correct, we replaced the affected sections by zeros. Transpiration increases the conductivity of the skin and so impedances becomes better. This leads to a better measurement of brain activity over longer periods, and thus could skew them. To flatten these long-term changes in electrode impedances, the signal was high-pass filtered with a cut-off frequency of 0.5 Hz. All steps of artefact correction have been applied to all 75 individual data sets.
	
	\item Grand-averaging or ERP calculation: The brain activity measured with an EEG is caused only to a small extent by the stimulus. A larger part is spontaneous activity and of a random nature. To enhance the signal to noise ratio (SNR), and thereby reveal the evoked event-related potentials (ERPs) across subjects, grand-averages for each electrode have been calculated by averaging the datasets of all subjects and trials \cite{Zanto.2005,Luck.2014,Woodman.2010}. This averaged data set was used to calculate synchronization between different electrodes and therefore between different brain regions.
	
	\item Frequency filter: The recorded EEG data was decomposed into common frequency bands: delta (0.5Hz - 3.5Hz), theta (3.5Hz - 7.5Hz), alpha (7.5Hz - 12.5Hz), beta (12.5Hz - 30Hz), and gamma (30Hz - 80Hz) using the finite impulse response (FIR) filter integrated into Matlab's EELab toolbox (v. 13.2.2b) resulting in 5 individual datasets.
\end{enumerate}

\subsubsection{Calculation of the synchronization}

Basically, synchronization between a pair of electrodes, and so between different brain regions, has been calculated for one second time windows by calculating the Pearson’s correlation coefficient r, the covariance between the neural activity of the one second time window of two electrodes, divided by the product of their standard deviations. This gives a value of $1 \geq r \leq -1$. For a completely in-phase synchronization of the signals r = 1, for a completely anti-phase synchronization of the signals r = -1, and for no synchronization r = 0 \cite{Davidova.2014,Nazemi.2018}.

By calculating the correlation coefficient for the successive time-windows over stimulus length, a course of synchronization between a pair of electrodes over time is obtained. This calculation was performed for all electrode pairs ($32 \times 31/2=496$). In order to determine the global synchronization of all brain regions represented by the individual electrodes, the curves of all electrode-pairs were averaged. This procedure was performed for all 5 filtered datasets.

Of particular interest is whether the local synchronization maxima match the times of the musical events. In order to determine the time points of the maxima, the areas in which the local maxima are searched for must first be defined. These are defined as the time of the musical events $\pm$ eight bars, shown in Tab. \ref{tab:maxima}. Eight bars correspond to a stimulus length of 15.36 sec (one beat = 60 sec / 125 bpm = 0.48 sec $\times$ 4 beats per bar $\times$ 8 bars = 15.36 sec). Since the time windows for synchronization is 1 sec, we will round time spans to 16 sec for calculation.

In addition to the maxima, the minima preceding the maxima are also interesting as comparison events, to see whether the strength of the global synchronization differs between the time points of the minima before a musical event and the time points of the musical event. In this way it can be determined whether the global synchronization has increased at the time point of a musical event. The time points of the minima are taken as in between the time points of the musical events.

\subsubsection{Statistical analysis of differences between local synchronization minima and musical events synchronization}

With a Paired Samples t Test the hypothesis can be tested, whether the synchronization strength at time points of musical events differ significantly from the synchronization strength at times points of local minima. The synchronization strength or values of all electrode pairs at the corresponding time points are used as synchronization values. In order to assess the meaning of a result, effect strengths are calculated. The effect strengths of the mean differences between groups can be determined by calculating a correlation coefficient r using the t value and the degrees of freedom. According to \cite{Cohen.1992}, r = 0.1 corresponds to a weak effect, r = .30 corresponds to a medium effect, and r = .50 corresponds to a strong effect.

\section{Results}

\subsection{Synchronization analysis}
\begin{figure}[h!]
		\includegraphics[width=0.5\textwidth]{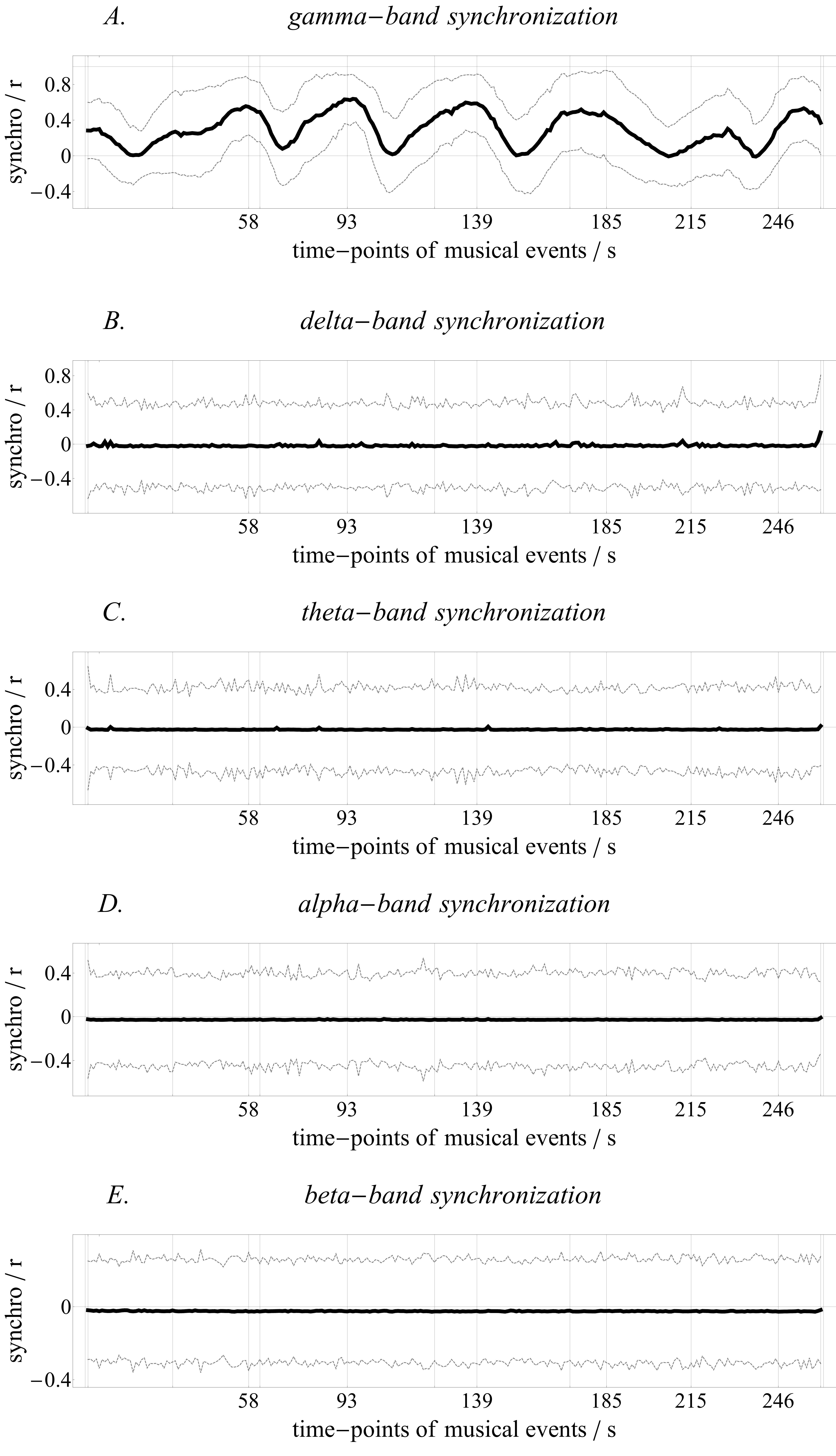}
		\caption{Averaged synchronization time series of all electrode pairs for different filtered datasets (black lines) supplemented by the standard deviation (dotted lines). The noted times mark the musical events. A) The averaged synchronization time series of all electrode-pairs in the gamma-band spectrum shows enhanced synchronization around time-points of musical events. B-E) The averaged synchronization time series of all electrode pairs in the filtered datasets in the delta-, theta-, alpha-, and beta-band spectrum show no such dynamics.}
	\label{fig:synchro_all}
\end{figure}

In Fig. \ref{fig:synchro_all} the time series of synchronization of all filtered datasets are shown. The solid vertical lines represent the boarders between the form parts of the song, as discussed in the stimulus section.

It can clearly be seen that local synchronization maxima in the gamma-band synchronization time series correspond to the musical events in that way that synchronization increases before a musical event occurs and decreases afterwards, while synchronization time series in all other filtered bands do not show any dynamics. The following results therefore refer to the averaged synchronization time series of all electrode pairs in the gamma-band. The local maxima are preluded by a rise and followed by a steeper decline in correlation. Tab. \ref{tab:maxima} shows the exact time points of the local synchronization maxima in the gamma-band synchronization time series corresponding to the musical events. The local synchronization maxima are very close to the time points of the musical events, but do not fit exactly.

\begin{table}[htbp]
\begin{tabular}{llr}
\hline
\begin{tabular}[c]{@{}l@{}}Musical event\\   / in sec\end{tabular} & \begin{tabular}[c]{@{}l@{}}Local maxima \\   time span / in sec\end{tabular} & \multicolumn{1}{l}{\begin{tabular}[c]{@{}l@{}}Local synchronization\\  maxima / in sec\end{tabular}} \\ \hline
58                                                                 & 042 - 074                                                                    & 57                                                                                                    \\ 
93                                                                 & 077 - 109                                                                    & 95                                                                                                    \\ 
139                                                                & 123 - 155                                                                    & 135                                                                                                   \\ 
185                                                                & 169 - 201                                                                    & 176                                                                                                   \\ 
215                                                                & 199 - 231                                                                    & 228                                                                                                   \\ 
246                                                                & 230 - 262                                                                    & 255                                                                                                   \\ \hline
\end{tabular}
\caption{Time points of local synchronization maxima in the gamma-band filtered EEG signal.}
	\label{tab:maxima}
\end{table}

Besides the time points of local maxima, local minima are interesting too, to compare the magnitude of synchronization between maxima and minima. Tab. \ref{tab:minima} shows the exact time points of local synchronization minima in the gamma-band, preceding a corresponding musical event.

\begin{table}[htbp]
\begin{tabular}{llr}
\hline
\begin{tabular}[c]{@{}l@{}}Musical event\\   / in sec\end{tabular} & \begin{tabular}[c]{@{}l@{}}Local minima \\   time span / in sec\end{tabular} & \multicolumn{1}{l}{\begin{tabular}[c]{@{}l@{}}Local synchronization\\  minima / in sec\end{tabular}} \\ \hline
58                                                                 & 001 - 057                                                                    & 17                                                                                                    \\ 
93                                                                 & 058 - 138                                                                    & 70                                                                                                    \\ 
139                                                                & 139 - 184                                                                    & 109                                                                                                   \\ 
185                                                                & 185 - 214                                                                    & 153                                                                                                   \\ 
215                                                                & 215 - 246                                                                    & 207                                                                                                   \\ 
246                                                                & 246 - 262                                                                    & 238                                                                                                   \\ \hline
\end{tabular}
\caption{Time points of local synchronization minima in the gamma-band filtered EEG signal.}
	\label{tab:minima}
\end{table}

\begin{table*}
\centering
\begin{tabular}{lllllllllll}
\hline
ME time & \multicolumn{3}{l}{Local maxima} & \multicolumn{3}{l}{Local minima} & \multicolumn{4}{l}{Paired Samples Test}                                                                                                   \\ \hline
        & Time      & Mean      & SD       & Time      & Mean      & SD       & t      & df   & \begin{tabular}[c]{@{}l@{}}Sig. \\ 2-tailed\end{tabular} & \begin{tabular}[c]{@{}l@{}}r according\\ to Cohen\end{tabular} \\
58      & 57        & 0.01      & 0.33     & 17        & 0.55      & 0.32     & -37.93 & 1023 & p\textless{}.001                                         & 0.76                                                           \\
93      & 95        & 0.08      & 0.41     & 70        & 0.63      & 0.28     & -36.42 & 1023 & p\textless{}.001                                         & 0.75                                                           \\
139     & 135       & 0.02      & 0.41     & 109       & 0.59      & 0.33     & -35.28 & 1023 & p\textless{}.001                                         & 0.74                                                           \\
185     & 176       & 0.00      & 0.01     & 153       & 0.46      & 0.02     & -27.81 & 1023 & p\textless{}.001                                         & 0.66                                                           \\
215     & 228       & -0.01     & 0.01     & 207       & 0.10      & 0.01     & -10.24 & 1023 & p\textless{}.001                                         & 0.30                                                           \\
246     & 255       & -0.00     & 0.36     & 238       & 0.32      & 0.39     & -21.70 & 1023 & p\textless{}.001                                         & 0.56                                                           \\ \hline
\end{tabular}
\caption{Results of paired t-test between local synchronization maxima and preceding local minima.}
	\label{tab:ttest}
\end{table*}

\subsection{Statistical analysis of differences between local synchronization minima and musical events synchronization}

To test if local synchronization maxima corresponding to the musical events and preceding local minima differ significantly, a Paired Samples t-Test between the synchronization values of all electrode pairs at these time points were performed, using IBM SPSS Statistics 24, as shown in Tab. \ref{tab:ttest}.

It can be shown that the mean differences between local synchronization maxima corresponding to the musical events and local synchronization minima preceding musical events are significant for all tested groups. The local synchronization maxima corresponding to the first three musical events (seconds 58, 93 and 139) differ with a very high effect, according to Cohens correlation coefficient, the lowest effect can be observed at the musical event at second 215.

In case of varying the 16-bars metrical structure, local maxima the 185th second is particularly noteworthy, since in the musical piece the drop of the bass sets in too late at this 185th second, in contrast to the 16-bars metrical structure the piece is composed of throughout. Relative to the expected beginning of the next musical form part, the bass starts too late. However, the curve showing the correlation reaches its local maximum after 16 bars, remaining at the same level until the drop starts, and does not decline until after. 

\section{Discussion}

	The results show that global neural synchronization between different brain regions in the gamma-band range increase before musical high-level events occur, and decreases afterwards. Paired Samples t-Test analysis shows significant differences between synchronization maxima corresponding to the defined musical events and preceding synchronization minima, with a strong effect for 5 of 6 tests.

Since the perception of the musical form is related to a a multiple of cortical processes, including timing and expectation as well as Gestalt formation and structured attention, it is hard to determine exactly what modulates large-scale synchronized neural activity while listening to a piece of music.

Since synchronization analysis is performed with a dataset grand-averaged over 25 subjects and 75 recordings in total, the synchronized neural activity must be phase-locked to the stimulus, and must be the same in all subjects. The shown synchronization can therefore be seen as a perceptual process and not as an individual experience of the subjects \cite{Zanto.2005}.

Since all subjects have experience with Tech-House music, and have therefore implicit knowledge about the structure of the genre \cite{Lerdahl.1990}, expectations about the high-level structure of the piece, the musical form, are generated by listening to a piece belonging to the genre \cite{Huron.2006,Huron.2012}. Since the form of this genre is pretty straight forward (16-bar structure), the time points of the musical events are predictable \cite{Friston.2013}. By counting bars it would be possible for a trained listener to predict the occurrence of a musical events very exactly. Since the task for subjects was just to listen, predictions made were probably not that exact, and by this it can be explained why local maxima differ slightly from time points calculated by the spectral centroid, but were at a maximum level around these time points.

On a neural level, the cognitive processes underlying the perception of musical form, expectation, and feature integration or Gestalt perception, are associated with large-scale neural synchronization \cite{Buhusi.2005,TallonBaudry.1999}. Besides that, attention modulates neural activity independent of the specific task. Several authors\cite{Nikolic.2013}. \cite{Jones.2002} formulated a Dynamic Attending Theory that attention related to music is not equally distributed, but rather periodic. In that context it is reasonable that attention is channeled towards the predictable musical events on the level of musical form, and therefore supports the cognitive and perceptual processes underlying the perception of musical form by stronger modulation of synchronized neural activity.
	
	\addtocounter{page}{2}
\bibliographystyle{abbrv}
\bibliography{MusForm_Arxiv}

\begin{thebibliography}{10}

\bibitem{Baars.2006}
B.~J. Baars.
\newblock Global workspace theory of consciousness: toward a cognitive
  neuroscience of human experience.
\newblock In S.~Laureys, editor, {\em The Boundaries of Consciousness}, volume
  150 of {\em Progress in Brain Research}, pages 45--53. {Elsevier
  professional}, s.l., 2006.

\bibitem{Bader.2013}
R.~Bader.
\newblock {\em Nonlinearities and Synchronization in Musical Acoustics and
  Music Psychology}.
\newblock {Springer Berlin Heidelberg}, Berlin, Heidelberg, 2013.

\bibitem{Bader.2018}
R.~Bader.
\newblock Cochlear spike synchronization and neuron coincidence detection
  model.
\newblock {\em Chaos (Woodbury, N.Y.)}, 28(2):023105, 2018.

\bibitem{Bader.2009}
R.~Bader, M.-K. Dietz, P.~Elvers, M.~Elias, and N.~Tolkien.
\newblock Foundation of a syllogistic music theory.
\newblock In R.~Bader, editor, {\em Musical Acoustics, Neurocognition and
  Psychology of Music / Musikalische Akustik, Neurokognition und
  Musikpsychologie}, Hamburger Jahrbuch f{\"u}r Musikwissenschaft, pages
  177--196. 2009.

\bibitem{Bhattacharya.2001}
J.~Bhattacharya, H.~Petsche, and E.~Pereda.
\newblock Long-range synchrony in the gamma band: Role in music perception.
\newblock {\em The Journal of Neuroscience}, 21(16):6329--6337, 2001.

\bibitem{Bregman.1990}
A.~S. Bregman.
\newblock {\em Auditory scene analysis : the perceptual organization of sound}.
\newblock {MIT Press}, Cambridge and Mass. [u.a.], 1990.

\bibitem{Buhusi.2005}
C.~V. Buhusi and W.~H. Meck.
\newblock What makes us tick? functional and neural mechanisms of interval
  timing.
\newblock {\em Nature reviews. Neuroscience}, 6(10):755--765, 2005.

\bibitem{Buhusi.2009}
C.~V. Buhusi and W.~H. Meck.
\newblock Relativity theory and time perception: single or multiple clocks?
\newblock {\em PloS one}, 4(7):e6268, 2009.

\bibitem{Burnham.2006}
S.~Burnham.
\newblock Form.
\newblock In T.~S. Christensen, editor, {\em The Cambridge history of Western
  music theory}, pages 880--906. {Cambridge Univ. Press}, Cambridge, 2006.

\bibitem{Cohen.1992}
J.~Cohen.
\newblock A power primer.
\newblock {\em Psychological bulletin}, 112(1):155--159, 1992.

\bibitem{Davidova.2014}
L.~Davidova, S.~{\'U}jv{\'a}ri, and Z.~N{\'e}da.
\newblock Sync or anti-sync -- dynamical pattern selection in coupled
  self-sustained oscillator systems.
\newblock {\em Journal of Physics: Conference Series}, 510:012009, 2014.

\bibitem{Dayan.1995}
P.~Dayan, G.~E. Hinton, R.~M. Neal, and R.~S. Zemel.
\newblock The helmholtz machine.
\newblock {\em Neural Computation}, 7(5):889--904, 1995.

\bibitem{Dehaene.2011b}
S.~Dehaene, J.-P. Changeux, and L.~Naccache.
\newblock The global neuronal workspace model of conscious access: From
  neuronal architectures to clinical applications.
\newblock In S.~Dehaene and Y.~Christen, editors, {\em Characterizing
  Consciousness: From Cognition to the Clinic?}, Research and Perspectives in
  Neurosciences, pages 55--84. {Springer-Verlag Berlin Heidelberg}, Berlin,
  Heidelberg, 2011.

\bibitem{Deliege.2014}
I.~Deli{\`e}ge and M.~M{\'e}len.
\newblock Cue abstraction in the representation musical form.
\newblock In I.~Deli{\`e}ge and J.~A. Sloboda, editors, {\em Perception and
  cognition of music}, pages 387--412. {Psychology Press}, Hove, 2014.

\bibitem{Deutsch.2013}
D.~Deutsch.
\newblock {\em The psychology of music}.
\newblock Academic Press series in cognition and perception. Academic, Oxford,
  3rd ed. edition, 2013.

\bibitem{Engel.2001}
A.~K. Engel, P.~Fries, and W.~Singer.
\newblock Dynamic predictions: oscillations and synchrony in top-down
  processing.
\newblock {\em Nature reviews. Neuroscience}, 2(10):704--716, 2001.

\bibitem{Engel.2001b}
A.~K. Engel and W.~Singer.
\newblock Temporal binding and the neural correlates of sensory awareness.
\newblock {\em Trends in Cognitive Sciences}, 5(1):16--25, 2001.

\bibitem{Freeman.2015}
W.~J. Freeman.
\newblock Mechanism and significance of global coherence in scalp eeg.
\newblock {\em Current Opinion in Neurobiology}, 31:199--205, 2015.

\bibitem{Freeman.2013}
W.~J. Freeman and R.~{Quian Quiroga}.
\newblock {\em Imaging brain function with EEG: Advanced temporal and spatial
  analysis of electroencephalographic signals}.
\newblock Springer, New York, 2013.

\bibitem{Fries.2009}
P.~Fries.
\newblock Neuronal gamma-band synchronization as a fundamental process in
  cortical computation.
\newblock {\em Annual review of neuroscience}, 32:209--224, 2009.

\bibitem{Friston.2013}
K.~Friston and {Friston Dimonic A.}
\newblock A free energy formulation of music generation and perception:
  Helmholtz revisited.
\newblock In R.~Bader, editor, {\em Sound - perception - performance}, Current
  Research in Systematic Musicology, pages 43--69. Springer, Cham [u.a.], 2013.

\bibitem{Friston.2006}
K.~Friston, J.~Kilner, and L.~Harrison.
\newblock A free energy principle for the brain.
\newblock {\em Journal of physiology, Paris}, 100(1-3):70--87, 2006.

\bibitem{Friston.2007}
K.~J. Friston and K.~E. Stephan.
\newblock Free-energy and the brain.
\newblock {\em Synthese}, 159(3):417--458, 2007.

\bibitem{Gibbon.1977}
J.~Gibbon.
\newblock Scalar expectancy theory and weber's law in animal timing.
\newblock {\em Psychological Review}, 84(3):279--325, 1977.

\bibitem{Gibbon.1984}
J.~Gibbon, R.~M. Church, and W.~H. Meck.
\newblock Scalar timing in memory.
\newblock {\em Annals of the New York Academy of Sciences}, 423:52--77, 1984.

\bibitem{Gray.1989}
C.~M. Gray and W.~Singer.
\newblock Stimulus-specific neuronal oscillations in orientation columns of cat
  visual cortex.
\newblock {\em Proceedings of the National Academy of Sciences of the United
  States of America}, 86(5):1698--1702, 1989.

\bibitem{GuevaraErra.2017}
R.~{Guevara Erra}, J.~L. {Perez Velazquez}, and M.~Rosenblum.
\newblock Neural synchronization from the perspective of non-linear dynamics.
\newblock {\em Frontiers in computational neuroscience}, 11:98, 2017.

\bibitem{Holson.1996}
R.~Holson, J.~F. Bowyer, P.~Clausing, and B.~Gough.
\newblock Methamphetamine-stimulated striatal dopamine release declines rapidly
  over time following microdialysis probe insertion.
\newblock {\em Brain Research}, 739(1-2):301--307, 1996.

\bibitem{Huron.2006}
D.~Huron.
\newblock {\em Sweet anticipation: Music and the psychology of expectation}.
\newblock A Bradford book. {MIT Press}, Cambridge, Mass., 2006.

\bibitem{Huron.2012}
D.~Huron and E.~H. Margulis.
\newblock Musical expectancy and thrills.
\newblock In P.~N. Juslin and J.~Sloboda, editors, {\em Handbook of music and
  emotion}, Series in affective science, pages 575--604. {Oxford University
  Press}, Oxford, 2012.

\bibitem{Ille.2002}
N.~Ille, P.~Berg, and M.~Scherg.
\newblock Artifact correction of the ongoing eeg using spatial filters based on
  artifact and brain signal topographies.
\newblock {\em Journal of clinical neurophysiology : official publication of
  the American Electroencephalographic Society}, 19(2):113--124, 2002.

\bibitem{Janata.1993}
P.~Janata and H.~Petsche.
\newblock Spectral analysis of the eeg as a tool for evaluating expectancy
  violations of musical contexts.
\newblock {\em Music Perception: An Interdisciplinary Journal}, 10(3):281--304,
  1993.

\bibitem{Jasper.1958}
H.~H. Jasper.
\newblock The ten-twenty electrode system of the international federation.
\newblock {\em Electroencephalography and Clinical Neurophysiology},
  10:371--375, 1958.

\bibitem{Jiruska.2013}
P.~Jiruska, M.~de~Curtis, J.~G.~R. Jefferys, C.~A. Schevon, S.~J. Schiff, and
  K.~Schindler.
\newblock Synchronization and desynchronization in epilepsy: controversies and
  hypotheses.
\newblock {\em The Journal of physiology}, 591(4):787--797, 2013.

\bibitem{Jones.2002}
M.~R. Jones, H.~Moynihan, N.~MacKenzie, and J.~Puente.
\newblock Temporal aspects of stimulus-driven attenting in dynamic arrays.
\newblock {\em Psychological Science}, 13(4):313--319, 2002.

\bibitem{Joris.1994}
P.~X. Joris, L.~H. Carney, P.~H. Smith, and T.~C. Yin.
\newblock Enhancement of neural synchronization in the anteroventral cochlear
  nucleus. i. responses to tones at the characteristic frequency.
\newblock {\em Journal of neurophysiology}, 71(3):1022--1036, 1994.

\bibitem{Joris.1994b}
P.~X. Joris, P.~H. Smith, and T.~C. Yin.
\newblock Enhancement of neural synchronization in the anteroventral cochlear
  nucleus. ii. responses in the tuning curve tail.
\newblock {\em Journal of neurophysiology}, 71(3):1037--1051, 1994.

\bibitem{Kelso.1997}
J.~A.~S. Kelso.
\newblock {\em Dynamic patterns: The self-organization of brain and behavior}.
\newblock A Bradford book. {MIT Press}, Cambridge, Mass., paperback ed.
  edition, 1997.

\bibitem{Koelsch.2013}
S.~Koelsch, M.~Rohrmeier, R.~Torrecuso, and S.~Jentschke.
\newblock Processing of hierarchical syntactic structure in music.
\newblock {\em Proceedings of the National Academy of Sciences of the United
  States of America}, 110(38):15443--15448, 2013.

\bibitem{Krumhansl.1979}
C.~L. Krumhansl and R.~N. Shepard.
\newblock Quantification of the hierarchy of tonal functions within a diatonic
  context.
\newblock {\em Journal of Experimental Psychology: Human Perception and
  Performance}, 5(4):579--594, 1979.

\bibitem{Kurth.1931}
E.~Kurth.
\newblock {\em Musikpsychologie}.
\newblock Hesse, Berlin, 1931.

\bibitem{Large.2015}
E.~W. Large, J.~A. Herrera, and M.~J. Velasco.
\newblock Neural networks for beat perception in musical rhythm.
\newblock {\em Frontiers in systems neuroscience}, 9:159, 2015.

\bibitem{Leman.1995}
M.~Leman.
\newblock {\em Music and schema theory : cognitive foundations of systematic
  musicology ; with 101 figures}.
\newblock Springer series in information sciences. Springer, Berlin [u.a.],
  1995.

\bibitem{Lerdahl.1990}
F.~Lerdahl and R.~Jackendoff.
\newblock {\em A generative theory of tonal music}.
\newblock The MIT Press series on cognitive theory and mental representation.
  {MIT Press}, Cambridge, Mass., 4. print edition, 1990.

\bibitem{Luck.2014}
S.~J. Luck.
\newblock {\em An introduction to the event-related potential technique}.
\newblock {MIT Press}, 2014.

\bibitem{Melloni.2007}
L.~Melloni, C.~Molina, M.~Pena, D.~Torres, W.~Singer, and E.~Rodriguez.
\newblock Synchronization of neural activity across cortical areas correlates
  with conscious perception.
\newblock {\em Journal of Neuroscience}, 27(11):2858--2865, 2007.

\bibitem{Meyer.1956}
L.~B. Meyer.
\newblock {\em Emotion and meaning in music}.
\newblock {Univ. of Chicago Press}, Chicago, 1956.

\bibitem{Mima.2001}
T.~Mima, T.~Oluwatimilehin, T.~Hiraoka, and M.~Hallet.
\newblock Transient interhemispheric neuronal synchronsy correlates with object
  recognition.
\newblock {\em The Journal of Neuroscience}, 21(11):3942--3948, 2001.

\bibitem{Nazemi.2018}
P.~S. Nazemi and Y.~Jamali.
\newblock On the influence of structural connectivity on the correlation
  patterns and network synchronization.
\newblock {\em Frontiers in computational neuroscience}, 12:105, 2018.

\bibitem{Neuhaus.2013}
C.~Neuhaus.
\newblock Processing musical form: Behavioural and neurocognitive approaches.
\newblock {\em Musicae Scientiae}, 17(1):109--127, 2013.

\bibitem{Nikolic.2013}
D.~Nikoli{\'c}, P.~Fries, and W.~Singer.
\newblock Gamma oscillations: precise temporal coordination without a
  metronome.
\newblock {\em Trends in Cognitive Sciences}, 17(2):54--55, 2013.

\bibitem{Nowak.2017}
A.~Nowak, R.~R. Vallacher, M.~Zochowski, and A.~Rychwalska.
\newblock Functional synchronization: The emergence of coordinated activity in
  human systems.
\newblock {\em Frontiers in psychology}, 8:945, 2017.

\bibitem{Owen.2019}
M.~Owen and M.~P. Guta.
\newblock Physically sufficient neural mechanisms of consciousness.
\newblock {\em Frontiers in systems neuroscience}, 13:24, 2019.

\bibitem{Palacios.2019}
E.~R. Palacios, T.~Isomura, T.~Parr, and K.~Friston.
\newblock The emergence of synchrony in networks of mutually inferring neurons.
\newblock {\em Scientific reports}, 9(1):6412, 2019.

\bibitem{Pastor.2012}
J.~Pastor, R.~G. de~Sola, and G.~J.
\newblock Hyper-synchronization, de-synchronization, synchronization and
  seizures.
\newblock In D.~Stevanovic, editor, {\em Automated Non-Invasive Identification
  and Localization of Focal Epileptic Activity by Exploiting Information
  Derived from Surface EEG Recordings}. {INTECH Open Access Publisher}, 2012.

\bibitem{Pearce.2012}
M.~T. Pearce and G.~A. Wiggins.
\newblock Auditory expectation: the information dynamics of music perception
  and cognition.
\newblock {\em Topics in cognitive science}, 4(4):625--652, 2012.

\bibitem{Perez.2017}
A.~P{\'e}rez, M.~Carreiras, and J.~A. Du{\~n}abeitia.
\newblock Brain-to-brain entrainment: Eeg interbrain synchronization while
  speaking and listening.
\newblock {\em Scientific reports}, 7(1):4190, 2017.

\bibitem{Petsche.1988}
H.~Petsche, K.~Linder, P.~Rappelsberger, and G.~Gruber.
\newblock The eeg: An adequate method to concretize brain processes elicited by
  music.
\newblock {\em Music Perception: An Interdisciplinary Journal}, 6(2):133--159,
  1988.

\bibitem{Reppert.2002}
S.~M. Reppert and D.~R. Weaver.
\newblock Coordination of circadian timing in mammals.
\newblock {\em Nature}, 418(6901):935--941, 2002.

\bibitem{Riemann.1976}
H.~Riemann.
\newblock {\em Pr{\"a}ludien und Studien gesammelte Aufs{\"a}tze zur Aesthetik,
  Theorie und Geschichte der Musik}.
\newblock Musikalische Studien. Kraus, Nendeln, reprint [d. ausg. 1895 u. 1900]
  edition, 1976.

\bibitem{Rodriguez.1999}
E.~Rodriguez, N.~George, J.~P. Lachaux, J.~Martinerie, B.~Renault, and F.~J.
  Varela.
\newblock Perception's shadow: long-distance synchronization of human brain
  activity.
\newblock {\em Nature}, 397(6718):430--433, 1999.

\bibitem{Rothfarb.2006}
L.~Rothfarb.
\newblock Energetics.
\newblock In T.~S. Christensen, editor, {\em The Cambridge history of Western
  music theory}, pages 927--955. {Cambridge Univ. Press}, Cambridge, 2006.

\bibitem{Salinas.2001}
E.~Salinas and T.~J. Sejnowski.
\newblock Correlated neuronal activity and the flow of neural information.
\newblock {\em Nature reviews. Neuroscience}, 2(8):539--550, 2001.

\bibitem{Schoenberg.1967}
A.~Schoenberg, editor.
\newblock {\em Fundamentals of musical composition}.
\newblock {Faber Faber}, London, 1967.

\bibitem{Seashore.1938}
C.~E. Seashore.
\newblock {\em Psychology of music}.
\newblock McGraw-Hill, New York, London, 1938.

\bibitem{Snoman.2009}
R.~Snoman.
\newblock {\em Dance music manual: Tools, toys and techniques. - Previous ed.:
  2004. - Accompanying CD-ROM includes audio examples in mp3 form. - Includes
  index}.
\newblock Focal, Amsterdam, 2nd ed. edition, 2009.

\bibitem{Snyder.2005}
J.~S. Snyder and E.~W. Large.
\newblock Gamma-band activity reflects the metric structure of rhythmic tone
  sequences.
\newblock {\em Brain research. Cognitive brain research}, 24(1):117--126, 2005.

\bibitem{Solberg.2016}
R.~T. Solberg and A.~R. Jensenius.
\newblock Pleasurable and intersubjectively embodied experiences of electronic
  dance music.
\newblock {\em Empirical Musicology Review}, 11(3-4):301--318, 2016.

\bibitem{Stumpf.1965}
C.~Stumpf.
\newblock {\em Tonpsychologie}.
\newblock Knuf, Hilversum, 1965.

\bibitem{Tallon.1995}
C.~Tallon, O.~Bertrand, P.~Bouchet, and J.~Pernier.
\newblock Gamma-range activity evoked by coherent visual stimuli in humans.
\newblock {\em European Journal of Neuroscience}, 7(6):1285--1291, 1995.

\bibitem{TallonBaudry.1999}
C.~Tallon-Baudry and O.~Bertrand.
\newblock Oscillatory gamma activity in humans and its role in object
  representation.
\newblock {\em Trends in Cognitive Sciences}, 3(4):151--162, 1999.

\bibitem{Thaut.2015}
M.~H. Thaut, G.~C. McIntosh, and V.~Hoemberg.
\newblock Neurobiological foundations of neurologic music therapy: rhythmic
  entrainment and the motor system.
\newblock {\em Frontiers in psychology}, 5, 2015.

\bibitem{Wagemans.2012}
J.~Wagemans, J.~H. Elder, M.~Kubovy, S.~E. Palmer, M.~A. Peterson, M.~Singh,
  and R.~von~der Heydt.
\newblock A century of gestalt psychology in visual perception: I. perceptual
  grouping and figure-ground organization.
\newblock {\em Psychological bulletin}, 138(6):1172--1217, 2012.

\bibitem{Wagemans.2012b}
J.~Wagemans, J.~Feldman, S.~Gepshtein, R.~Kimchi, J.~R. Pomerantz, P.~A. {van
  der Helm}, and C.~{van Leeuwen}.
\newblock A century of gestalt psychology in visual perception: Ii. conceptual
  and theoretical foundations.
\newblock {\em Psychological bulletin}, 138(6):1218--1252, 2012.

\bibitem{Witek.2016}
M.~A.~G. Witek and P.~Vuust.
\newblock Comment on solberg and jesenius: The temporal dynamics of embodied
  pleasure in music.
\newblock {\em Empirical Musicology Review}, 11(3-4):324--329, 2016.

\bibitem{Womelsdorf.2007}
T.~Womelsdorf and P.~Fries.
\newblock The role of neuronal synchronization in selective attention.
\newblock {\em Current Opinion in Neurobiology}, 17(2):154--160, 2007.

\bibitem{Woodman.2010}
G.~F. Woodman.
\newblock A brief introduction to the use of event-related potentials in
  studies of perception and attention.
\newblock {\em Attention, perception {\&} psychophysics}, 72(8):2031--2046,
  2010.

\bibitem{Zanto.2005}
T.~P. Zanto, E.~W. Large, A.~Fuchs, and {Kelso, Kelso, J.A. Scott}.
\newblock Gamma-band responses to perturbed auditory sequences: Evidence for
  synchronization of perceptual processes.
\newblock {\em Music Perception: An Interdisciplinary Journal}, 22(3):531--547,
  2005.

\end{thebibliography}

\end{document}